\documentclass[twocolumn,showpacs,preprintnumbers,amsmath,amssymb,epsfig,widetext]{revtex4}
\usepackage{graphicx}
\usepackage{dcolumn}
\usepackage{bm}
\usepackage{epsfig}
\usepackage{color}

\def\fun#1#2{\lower3.6pt\vbox{\baselineskip0pt\lineskip.9pt
  \ialign{$\mathsurround=0pt#1\hfil##\hfil$\crcr#2\crcr\sim\crcr}}}
\def\simgt{\mathrel{\lower0.6ex\hbox{$\buildrel {\textstyle >}
 \over {\scriptstyle \sim}$}}}
\def\simlt{\mathrel{\lower0.6ex\hbox{$\buildrel {\textstyle <}
 \over {\scriptstyle \sim}$}}}

\input epsf

\newcommand{\hompc}{\,h\,{\rm Mpc}^{-1}}

\def\be{\begin{equation}}
\def\ee{\end{equation}}
\def\ba{\begin{eqnarray}}
\def\ea{\end{eqnarray}}

\def\nn{\nonumber}

\begin{document}

\preprint{}
\title{
Cosmology with anisotropic galaxy clustering from the 
combination of power spectrum and bispectrum}
\author{Yong-Seon Song$^1$, Atsushi Taruya$^{2,3}$, Akira Oka$^4$}
\affiliation{
$^1$Korea Astronomy and Space Science Institute, Daejeon 305-348, R. Korea and\\
$^2$Yukawa Institute for Theoretical Physics, Kyoto University, Kyoto 606-8502, Japan\\
$^3$Kavli Institute for the Physics and Mathematics of the Universe, Todai Institutes for Advanced Study, the University of Tokyo, Kashiwa, Chiba 277-8583, Japan\\
$^4$ Department of Physics, University of Tokyo, Tokyo 113-0033, Japan\\}
\date{\today}
\begin{abstract}
The apparent anisotropies of the galaxy clustering in observable redshift space provide a unique opportunity to simultaneously probe cosmic expansion and gravity on cosmological scales via the Alcock--Paczynski effect and redshift-space distortions. While the improved theoretical models have been proposed and developed to describe the apparent anisotropic clustering at weakly non-linear scales, the applicability of these models is still limited in the presence of the non--perturbative smearing effect caused by the randomness of the relative velocities. 
Although the cosmological constraint from the anisotropic clustering will be improved with a more elaborate theoretical model, here we consider an alternative approach using the statistical power of both the power spectrum and bispectrum at large scales. Based on the Fisher matrix analysis, we estimate the benefit of combining the power spectrum and bispectrum, finding 
that for the future spectroscopy survey DESI (Dark Energy Spectroscopy Instrument), the constraints on the cosmic expansion and growth of structure will be improved by a factor of two. This approach compensates for the loss of constraining power, using the power spectrum alone, due to the randomness of the relative velocities. 
\end{abstract}
\preprint{YITP-15-7}

\maketitle
\section{Introduction}
\label{sec:intro}

In our current understanding of the universe, an unknown substance called dark matter dominates over the standard model particles at the present epoch. Despite many theoretical and observational efforts, the origin of dark matter is not yet clarified. Also, the existence of dark energy, which is supposed to drive the cosmic acceleration, indicates our incomplete understanding of the gravity on cosmological scales~\cite{Riess:1998cb,Perlmutter:1998np}. It may imply modifications to Einstein's  theory of General Relativity.  A further insight into the origin and nature of dark energy or validity of general relativity is essential, and this is one of the primary goals in next-generation cosmology.

The large-scale structure offers an opportunity to probe these issues by looking at the anisotropic galaxy clustering in redshift space~\cite{Song:2008qt,Wang:2007ht,Nesseris:2007pa,White:2008jy,Percival:2008sh}. The observed galaxy distribution via the spectroscopic measurements is apparently distorted due to the peculiar velocity of galaxies along the line-of-sight direction, referred to as the redshift-space distortions (RSD). While the RSD complicates the interpretation of the small-scale galaxy clustering, on large scales, the strength of anisotropies is simply characterized by the linear growth rate $f=d\ln \delta_{\rm m}/d\ln a$ \cite{Kaiser:1987qv}, providing us a unique opportunity to probe the growth of structure, where $\delta_{\rm m}$ and $a$ are the linear density field and scale factor of the Universe, respectively. On the other hand, the large-scale galaxy clustering data imprints a fossil record of the primeval baryon-photon fluid around the last-scattering surface, known as the baryon acoustic oscillations (BAO)~\cite{Seo:2003pu,Blake:2003rh,Gaztanaga:2008xz}. The characteristic scale of the BAO can be used as a standard ruler, which enables us to determine the geometric distances of high-$z$ galaxies with a greater precision. The key point to determine the geometric distances is to measure the clustering anisotropies over the BAO scales. Notice the fact that the anisotropies of the clustering pattern also arises from the apparent mismatch of the underlying cosmological model when we convert the redshift and angular position of each galaxy to the co-moving radial and transverse distances. This is the so-called Alcock-Paczynski (A-P) effect~\cite{Alcock:1979mp}, and with a prior knowledge of the characteristic scale of the BAO, the Hubble parameter $H(z)$ and angular diameter distance $D_A(z)$ of the high-$z$ galaxies can be separately measured. Thus, the anisotropic galaxy clustering can serve as a dual cosmological probe from which we can explore the origin of cosmic acceleration from the viewpoint of both dark energy and modification of gravity.

In order to simultaneously extract information on both the growth of structure and cosmic expansion, a detailed theoretical model of the anisotropic clustering is crucial. While the scales of our interest are rather close to the linear regime of structure formation, there appear small but non-negligible nonlinear systematics that must be corrected or subtracted, and these including the nonlinear gravitational clustering. This is one of the main reasons why perturbation theory has become very popular recently as a theoretical template of the power spectrum or correlation function beyond the linear theory (e.g., \cite{Jeong:2006xd,Valageas:2007,Crocce:2007dt,Taruya:2008,Matsubara:2007wj,Pietroni:2008,Bernardeau:2008,Taruya:2009,Taruya:2012}). Still, however,  perturbation theory has its own limitations, and we cannot apply it to the small scales beyond the weakly nonlinear regime. Furthermore, the RSD described by the nonlinear mapping from the real to redshift spaces, gives an additional complication that leads to the non-trivial cross talk between small- and large-scale clustering. As a result, even at large scales, the clustering amplitude is significantly reduced along the line of sight, known as the Finger-of-God effect. While several treatments, based on perturbation theory, have been proposed in order to precisely model the nonlinear RSD (e.g., \cite{Fisher:1994ks,Scoccimarro:2004tg,Taruya:2010mx,Jennings:2010uv,Reid:2011ar,Okumura:2011pb,Kwan:2011hr,Zhang:2012yt,Taruya:2013}), the Finger-of-God effect, being most likely ascribed to the virialized random motion the galaxies inside halos, prevents us from a detailed modeling within perturbation theory. Hence, a phenomenological description characterizing the Finger-of-God suppression needs to be introduced, and in order to avoid any unwanted systematics, we have to conservatively restrict the cosmological data analysis to the larger scales, $k\lesssim0.1\hompc$, for instance. This, needless to say, significantly reduces the statistical power to constrain dark energy or to test general relativity. 

In the persuit of extracting maximal cosmological information from the 
anisotropic galaxy clustering data, 
a simple but potentially powerful approach is to make use of the benefit of 
combining both the power spectrum and bispectrum on large scales.   
Although the initial condition for perturbations is supposed to be nearly 
Gaussian, non--vanishing bisepctrum naturally arises from the nonlinear mode
coupling through the late-time gravitational evolution. At the weakly nonlinear regime, the bispectrum still contains statistical information similar to the 
power spectrum, and thus the BAO feature should be clearly manifest. Hence, 
the bispectrum in redshift space, as an actual observable, 
can be used as an alternative tool to constrain the geometric 
distances and growth of structure via the A-P effect and 
RSD. Thus, in combining the bispectrum and the power spectrum, we expect a substantial on the resultant cosmological constraints. Indeed, 
in the context of cosmology with galaxy redshift surveys, 
the benefit of using the bispectrum in cosmological data analysis has already been studied
in several work (e.g., \cite{Fry:1992vr,Sefusatti:2006,Greig:2013,Gil-Marin:2014sta,Gil-Marin:2014baa}). Among these, Sefusatti, Crocce, Pueblas $\&$ Scoccimarro~\cite{Sefusatti:2006} focused on 
the angle-averaged bispectrum in redshift space, and discussed a potential power of bispectrum to constrain multiple cosmological parameters, ignoring A-P effect. On the other hand, taking fully account of both the A-P effect and RSD, 
Greig, Komatsu $\&$ Whyithe~\cite{Greig:2013} considered cosmology with Ly$\alpha$ emitting galaxies, and specifically studied the impact of radiative transfer effects on the observed clustering of Ly$\alpha$ emitting galaxies. It has been shown that the bispectrum is helpful to distinguish between gravitational and non-gravitational effects, thus breaking the parameter degeneracies.

In this paper, we present the combined results of the power spectrum and bispectrum to constrain the geometric distances and growth of structure. The present 
paper is partly similar to~\cite{Greig:2013}, but is rather different in several aspects. To be specific, we consider a future accessible spectroscopic survey like Dark Energy Spectroscopy Instrument (hereafter DESI), which would be the best suited to probe the cosmic acceleration around $z\sim1$. Considering this survey setup, we discuss the impact of the Finger-of-God effect on the estimation of cosmological parameters, which has not been considered in ~\cite{Greig:2013}. We show that while the uncertainty of the Finger-of-God effect in power spectrum is mostly degenerate with the coherent motion as a probe of the growth of structure, the combination of power spectrum and bispectrum breaks this degeneracy, thus improving the measurement accuracy of the coherent motion by a factor of two. As for the constraints on geometric distances, substantial improvement is found, consistent with previous works.  Furthermore, the role of the non-vanishing cross covariance between power spectrum and bispectrum is studied, and the relative impact in 
estimating statistical errors is quantified, finding that the influence of cross covariance is small enough in the weakly nonlinear regime.

This paper is organized as follows. In Sec.~\ref{sec:constraints}, 
the basic setup of the forecast analysis is presented. Based on 
perturbation theory, theoretical models for power spectrum and bispectrum 
are given, taking account of both the A-P effect and 
RSD. Then, the basic formalism for  
Fisher matrix analysis is described, including the cross covariance
between power spectrum and bispectrum. Sec.~\ref{sec:results} presents 
the main results for Fisher matrix analysis. After comparing 
the signal-to-noise ratios of power spectrum with those of bispectrum, 
we show the expected constraints on geometric distances ($D_A$ and $H$) 
and growth of structure $f$ in DESI-like experiments. 
The impact of Finger-of-God effect is discussed 
in detail, and role of the bispectrum is clarified. 
Finally, Sec.~\ref{sec:summary} is devoted to the summary and discussion.

\section{Constraints on cosmology using power spectrum and bispectrum}
\label{sec:constraints}

\subsection{Power spectrum}
\label{subsec:power_spectrum}

On large scales of our interest,  
the density and velocity fields are basically 
the small perturbations to the homogenous background. 
When the higher-order contributions are ignorable, 
the power spectrum is simply described by the linear theory, and 
in redshift space, it is given by
\ba\label{eq:kaiser}
\tilde{P}^{\rm lin} (\vec{k}) = Z_1^2(\vec{k})P(k) \,,
\ea
where $Z_1(\vec{k})$ is defined by,
\ba
Z_1(\vec{k})\equiv b+f\mu^2\,.
\ea
The linear galaxy bias, denoted by $b$, represents the enhancement of 
the clustering amplitude relative to the mass density field $\delta_{\rm m}$. The function $f$ is defined by the logarithmic derivative of the linear density field with respect to the logarithm of scale factor, 
i.e., $f=d\ln\delta_{\rm m}/d\ln a$. The directional vector $\vec{k}$ 
is decomposed into $(k,\mu)$ where $\mu$ denotes the cosine of angle to the line of sight.

In practice, the applicability of 
the linear theory expression in Eq.~(\ref{eq:kaiser})
is restricted to a narrow range of scales. 
This is because the mapping of statistical quantities from 
real to redshift space is intrinsically nonlinear. Even at large scales, 
higher-order contributions to the mapping formula are not negligible. 
For this reason, there have been several improved models of RSD that have been proposed that add correction terms to the Kaiser formula in Eq.~(\ref{eq:kaiser}). 
Taking account of those effects is thus crucial and essential 
for an unbiased parameter estimation in the practical data analysis.  
On the other hand, the estimation of statistical 
errors is not usually much dependent on those elaborate 
factorized formulations, because the statistical error of each 
parameter mainly comes from the measurement uncertainties, 
including the cosmic variance and shot noise.  Unless 
a significant contribution of higher-order corrections arises, 
the structure of parameter degeneracies will remain unchanged. 
Hence, in this paper, we do not consider such higher-order corrections.

Nevertheless, the suppression of clustering amplitude 
due to the random motion is known as a non--perturbative effect,
which significantly affects the power spectrum even at large scales, 
and should be accounted in the basic formulation of Eq.~(\ref{eq:kaiser}). 
Here, we assume that this FoG 
is given as a factorized form, and multiplied as (e.g., \cite{Scoccimarro:2004tg,PeacockDodds:1994,Park:1994fa,Ballinger:1996cd,Magira:199bn}):
\ba\label{eq:fullpks}
\tilde{P} (\vec{k}) = D^P_{\rm FoG}(\vec{k})\tilde{P}^{\rm lin} (\vec{k}) \,.
\ea
The $D^P_{\rm FoG}(\vec{k})$ is given by the Gaussian form as,
\ba\label{eq:fogpower}
D^P_{\rm FoG}(\vec{k}) = {\rm exp}\left[-\left(k\mu\sigma_p\right)^2\right]\,,
\label{eq:FoG_power}
\ea
where $\sigma_p$ denotes the dispersion of the one-point PDF of the velocity in one-dimension. Note that at smaller scales, 
the virial motion of galaxies inside a cluster of galaxies 
also leads to a suppression of the power spectrum in redshift space. 
When $(k\mu\sigma_p)^2 \ll 1$, the leading order term of Eq.~(\ref{eq:fullpks}) is dominant over all other higher orders, and the estimated errors are immune from the exact functional form of Eq.~(\ref{eq:fullpks}). The linear $\sigma_p$ is used for the fiducial value.

\subsection{Bispectrum}

While the initial condition for perturbations is assumed to be Gaussian, 
gravitational evolution naturally induces mode-mode coupling, 
giving rise to 
the non-vanishing bispectrum. Furthermore, coupled with galaxy bias and 
RSD, the bispectrum in redshift space becomes rather 
complicated. The resultant leading-order expression for the bispectrum~(e.g., \cite{Fry:1992vr}), valid at 
large scales, is given by,
\ba\label{eq:bilinear}
\tilde{B}^{\rm PT} (\vec{k}_1,\vec{k}_2,\vec{k}_3)&=&2\Big[Z_2(\vec{k}_1,\vec{k}_2)Z_1(\vec{k}_1)Z_1(\vec{k}_2)P(k_1)P(k_2)\,. \nn\\
&+&{\rm cyclic}\,\Big]
\ea
The kernel $Z_2$ is defined as,
\ba
Z_2(\vec{k}_i,\vec{k}_j)&\equiv&\frac{b_2}{2}+bF_2(\vec{k}_i,\vec{k}_j)
+f\mu_{ij}^2G_2(\vec{k}_i,\vec{k}_j)\\
&+&\frac{f\mu_{ij}k_{ij}}{2}\left[\frac{\mu_i}{k_i}(b+f\mu_j^2) + \frac{\mu_j}{k_j}(b+f\mu_i^2)\right]\,,\nn
\ea
where we define $\mu_i=(\vec{k}_i\cdot\hat{z})/k_i$, 
$\vec{k}_{ij}=\vec{k}_i+\vec{k}_j$, 
$\mu_{ij}=(\vec{k}_{ij}\cdot\hat{z})/k_{ij}$, with $\hat{z}$ being the 
line-of-sight unit vector. 
Here, we incorporate the uncertainty of the 
nonlinear galaxy bias characterized by $b_2$ into the kernel $Z_2$, 
adopting the local bias prescription (e.g., \cite{Fry:1992vr}), 
i.e., $\delta_{\rm g}=b\,\delta_{\rm m}+(b_2/2)\,\delta_{\rm m}^2+\cdots$. In the 
above, the functions $F_2$ and $G_2$ are the 
standard PT kernel in real space, given by,
\ba
F_2(\vec{k}_i,\vec{k}_j)&=&\frac{5}{7} + \frac{\eta_{ij}}{2} \left(\frac{k_i}{k_j}+\frac{k_j}{k_i}\right) + \frac{2}{7} \eta_{ij}^2 \\
G_2(\vec{k}_i,\vec{k}_j)&=&\frac{3}{7} + \frac{\eta_{ij}}{2} \left(\frac{k_i}{k_j}+\frac{k_j}{k_i}\right) + \frac{4}{7} \eta_{ij}^2\,.
\ea
with $\eta_{ij}=(\vec{k}_i\cdot\vec{k}_j)/(k_ik_j)$. Note that 
the configuration of bispectrum satisfies the triangular condition, 
which is expressed by the directional vector constraint,
\ba\label{eq:biconfig}
\vec{k}_1+\vec{k}_2+\vec{k}_3 = 0 \,.
\ea

In contrast to the redshift-space power spectrum, 
the influence of nonlinear RSD on Eq.~(\ref{eq:bilinear}) is not yet 
fully understood and studied in detail. 
Although it deserves further investigation, 
we can make the best guess on the possible damping effect 
due to the random motion of galaxy. The FoG effect in the bispectrum 
is assumed to be Gaussian as \cite{Scoccimarro:1999},
\ba\label{eq:fogbi}
D^B_{\rm FoG}(\vec{k}_1,\vec{k}_2,\vec{k}_3) = {\rm exp}\left[-(k^2_1\mu^2_1+k^2_2\mu^2_2 +k^2_3\mu^2_3) \sigma_p^2\right].
\label{eq:FoG_bispec}
\ea
Then the observed bispectrum is given by,
\ba\label{eq:bifull}
\tilde{B} (\vec{k}_1,\vec{k}_2,\vec{k}_3)=D^B_{\rm FoG}(\vec{k}_1,\vec{k}_2,\vec{k}_3)\tilde{B}^{\rm PT} (\vec{k}_1,\vec{k}_2,\vec{k}_3)\,.
\ea
Again, when $(k\mu\sigma_p)^2 \ll 1$, the detailed functional form of $D^B_{\rm FoG}$ is not important for our estimation.

\subsection{Alcock--Paczynski test}

In addition to the anisotropies induced by the RSD, the 
observed galaxy clustering also exhibits anisotropies through the 
Alcock-Paczynski (A-P) effect. This can happen if the background 
expansion of the real universe differs from the fiducial cosmology 
used to  convert the redshift and angular position of each galaxy 
to the co-moving radial and transverse distances.

While this effect leads to the modulation in the shape and amplitude of the power spectrum and bispectrum, if the shape of these quantities is a priori known, it offers a unique opportunity to measure 
the angular diameter distance $D_A(z)$ and Hubble parameter $H(z)$ of 
distant galaxies at redshift $z$ using the 
characteristic shape of the galaxy clustering in both the radial and transverse directions. 
Furthermore, providing information on the evolution of density and velocity 
fields, the two types of apparent anisotropies (i.e., RSD and A-P effects) 
become distinguishable, and the geometric distances $D_A$ and $H$ can be  
separately and accurately determined. 
This is indeed possible if we know at least 
the broadband shape of spectrum. In other words, given a accurate theoretical 
template which describes the broadband shape of the power spectrum 
and bispectrum, 
the simultaneous constraints on the geometric distances and growth of 
structure are made possible. We dub this method as a broadband A-P test~\cite{Song:2013ejh}.

The anisotropies in the power spectrum caused by the A-P effect are modeled 
as follows.  Denoting the true power spectrum by $\tilde{P}$ , the 
observed power spectrum becomes
\ba
\tilde{P}^{\rm obs} (k,\mu) = \left(\frac{H^{\rm true}}{H^{\rm fid}}\right)  \left(\frac{D_A^{\rm fid}}{D_A^{\rm true}}\right)^2 \tilde{P} (q,\nu)\,,
\label{eq:Pk_obs}
\ea
where $(k,\mu)$ denotes the fiducial coordinates for the underlying cosmological model, and $(q,\nu)$ represents the coordinates in the {\it true} cosmology.

The A-P effect for bispectrum is also modeled in a similar way, 
and the resultant shape of bispectrum depends on 
five parameters, i.e.,  $(k_1,k_2,k_3,\mu_1,\mu_2)$. The observed 
bispectrum is thus related to the true one given in Eq.~(\ref{eq:bifull}) through,
\ba
\tilde{B}^{\rm obs}(k_1,k_2,k_3,\mu_1,\mu_2)&=&\left(\frac{H^{\rm true}}{H^{\rm fid}}\right)^2  \left(\frac{D_A^{\rm fid}}{D_A^{\rm true}}\right)^4 \nn\\
&\times& \tilde{B}(q_1,q_2,q_3,\nu_1,\nu_2)\,.
\label{eq:Bk_obs}
\ea
The relations between two coordinates are give by,
\ba
q_i=\alpha(\mu_i)k_i\,,
\ea
and
\ba
\nu_i=\frac{\mu_i}{\alpha(\mu_i)}\frac{H^{\rm true}}{H^{\rm fid}}\,,
\ea
where $\alpha(\mu_i)$ is defined by,
\ba
\alpha(\mu_i)\equiv\left\{\left(\frac{D_A^{\rm fid}}{D_A^{\rm true}}\right)^2+\left[\left(\frac{H^{\rm true}}{H^{\rm fid}}\right)^2-\left(\frac{D_A^{\rm fid}}{D_A^{\rm true}}\right)^2\right]\mu_i^2\right\}^{1/2}\,.\nn
\ea
The cosine of angle between two vectors, 
$\nu_{ij}=(\vec{q}_i\cdot\vec{q}_j)/(q_iq_j)$, is given by,
\ba
\nu_{ij}&=&\left(\frac{D_A^{\rm fid}}{D_A^{\rm true}}\right)^2\frac{\eta_{ij}}{\alpha(\mu_i)\alpha(\mu_j)}\nn\\
&+&\left[\left(\frac{H^{\rm true}}{H^{\rm fid}}\right)^2-\left(\frac{D_A^{\rm fid}}{D_A^{\rm true}}\right)^2\right]\frac{\mu_{i}\mu_{j}}{\alpha(\mu_i)\alpha(\mu_j)}\,.
\ea
Here, we define $\eta_{ij}=(\vec{k}_i\cdot\vec{k}_j)/(k_ik_j)$.

\subsection{Covariance matrix and Fisher matrix analysis}

In this paper, to elucidate the potential power of the bispectrum to 
constrain cosmology,  we shall specifically consider DESI as a representative future galaxy 
survey, and proceed to the Fisher matrix analysis. 
The primary science goal of the DESI experiment is to clarify the 
nature of dark energy and/or gravity through the A-P and RSD effects,  
and starting in 2018, it will obtain optical spectra for tens of millions of galaxies and quasars, constructing a 3-dimensional map spanning the nearby universe to 10 billion light years. DESI will be conducted on the Mayall 4-meter telescope at Kitt Peak National Observatory. It is supported by the Department of Energy Office of Science to perform Stage IV dark energy measurement 
(see \cite{DETF:2006} for definition of Stage IV-class survey). 
The expected number density of the galaxies 
in terms of the co-moving volume is summarized in Table~\ref{tab:DESI}.

\begin{table}[th]
\begin{ruledtabular}
\begin{tabular}{c|c|c}
  $z$ & $n_{g}\,[h^3{\rm Mpc}^{-3}]$ & $V_{\rm survey}\,[h^{-3}\,{\rm Gpc}^3]$\\
  \hline
  0.6--0.8 & $1.2\times 10^{-3}$  & 5.3\\
  0.8--1.0 & $1.1\times 10^{-3}$  & 7.0\\
  1.0--1.2 & $5.4\times 10^{-4}$  & 8.3\\
  1.2--1.4 & $3.3\times 10^{-4}$  & 9.4\\
  1.4--1.6 & $1.5\times 10^{-4}$  & 10.1\\
  1.6--1.8 & $5.0\times 10^{-5}$  & 10.6\\
\end{tabular}
  \caption{\label{tab:DESI}
The expected number density of galaxies $n_g$ and 
survey volume $V_{\rm survey}$ at each redshift bin used in the Fisher matrix 
analysis. These specific values are taken from those assumed 
in the DESI experiment. }
\end{ruledtabular}
\end{table}

In order to compute the Fisher matrix, the error covariance of 
the power spectrum and bispectrum needs to be evaluated. To simplify the 
analysis, we will ignore the off-diagonal components of 
the covariance matrices arising mainly from the nonlinear mode coupling. 
This would certainly 
lead to an optimistic estimation of the parameter constraints, however, 
it has been shown that in the case of power spectrum,  
  the non-Gaussian error contribution 
  to the off-diagonal components are small enough, and 
  the diagonal components of the covariance matrix 
  can be approximately described by the simple Gaussian contribution 
  \cite{Takahashi:2009}. 
  This would be true as long as we consider 
  the quasi-linear scales at moderately high redshift (say, $z\gtrsim1$ and 
  $k\lesssim0.15\,h$\,Mpc$^{-1}$). The parameter estimation study with non-Gaussian covariance further revealed that 
  the size of the constraints on each parameter is not drastically changed 
  if we consider the multiple parameter estimation 
  (\cite{Takahashi:2011,Ngan:2012}, see also \cite{Takada:2009}).  
  
  The Gaussian contribution to the covariance matrix 
  for power spectrum is given by,
\ba
C_{\rm PP}=\frac{1}{N_P}\left[Z_1^2(k,\mu) P(k) + \frac{1}{n_g}\right]^2\,,
\ea
where $N_P$ is given by,
\ba
N_P=\frac{V_{\rm survey}}{2(2\pi)^2} k^2\Delta k\Delta\mu\,.
\ea
Here the survey volume is derived by calculating the co-moving shell in each redshift bin, and multiply by fractional factor from DESI $f_{\rm sky}=14,000\,\,{\rm deg}^2$. 
The Gaussian contribution to the covariance matrix for bispectrum is 
expressed as,
\ba
&&C_{\rm B_{\alpha}B_{\beta}}=\delta_{\alpha\beta}s_B\frac{V_{\rm survey}}{N_B}
\left[Z_1^2(k_1,\mu_1) P(k_1) + \frac{1}{n_g}\right]\\
&&\quad\times\left[Z_1^2(k_2,\mu_2) P(k_2) + \frac{1}{n_g}\right]
\left[Z_1^2(k_3,\mu_3) P(k_3) + \frac{1}{n_g}\right],\nn
\ea
where $N_B$ is given by,
\ba
N_B=2\pi k_1k_2k_3(\Delta k)^3(\Delta\mu)^2
\left[\frac{V_{\rm survey}}{(2\pi)^3}\right]^2\,.
\ea
The pre--factor $s_B$ is set to $6$ for equilateral, $2$ 
for isosceles, and $1$ for general triangular configurations. 
Finally, the cross-covariance matrix between power spectrum and bispectrum
is given by
\ba\label{eq:crosscov}
&&C_{\rm PB} = \frac{s_B}{N_P}
\left(\delta^D_{\vec{k}\vec{k}_1}+\delta^D_{\vec{k}\vec{k}_2}+\delta^D_{\vec{k}\vec{k}_3}\right) \left[Z_1^2 P(k) + n_g^{-1}\right]  \\
&&\times\left\{\frac{2Z_2\left[Z_1^2 P(k_1) + n_g^{-1}\right] \left[Z_1^2P(k_2) + n_g^{-1}\right]}{Z_1(k_1,\mu_1)Z_1(k_2,\mu_2)} + \rm cyclic \right\} \nn
\ea

Notice that while the covariance matrices, 
$C_{\rm PP}$ and $C_{\rm BB}$, 
can be expressed as a diagonal form, the full matrix of $C$ combining both power spectrum and 
bispectrum is no longer diagonal 
even in the Gaussian case. This generally requires a complicated
matrix algebra, however, the covariance matrix $C$ in our case can be 
expressed in a block-diagonal form as,
\ba
C=
\left(\begin{array}{cc}
C_{\rm PP} & C_{\rm PB} \\
C_{\rm BP} &  C_{\rm BB} \\
\end{array}\right)\,.
\ea
Then the inverse matrix $C^{-1}$ is expressed as, 
\ba
C^{-1}=\left(\begin{array}{cc}
M \,\, &\,\, -M\,C_{\rm PB} C_{\rm BB}^{-1} \\
-C_{\rm BB}^{-1}C_{\rm BP}M \,\,& \,\,C_{\rm BB}^{-1}+C_{\rm BB}^{-1}C_{\rm BP}MC_{\rm PB}C_{\rm BB}^{-1}\\
\end{array}\right)\,
\nonumber
\ea
with the matrix $M$ given by $M\equiv (C_{\rm PP}-C_{\rm PB}C_{\rm BB}^{-1}C_{\rm BP})^{-1}$. Note that with the Woodbury formula 
$M=C_{\rm PP}^{-1}-C_{\rm PP}^{-1}C_{\rm PB}(C_{\rm BB}^{-1}+C_{\rm BP}C_{\rm PP}^{-1}C_{\rm PB})^{-1}C_{\rm BP}C_{\rm PP}^{-1}$, we easily verify that $C\,C^{-1}=C^{-1}\,C=I$.

With the full covariance matrix given above, the Fisher matrix combining 
the power spectrum and bispectrum becomes
\ba\label{eq:fullfisher}
F_{\alpha\beta}&=&
\sum_{\vec{k}}\sum_{\vec{k}_1,\vec{k}_2,\vec{k}_3} 
\frac{\partial ~^{t}\vec{S}^{\rm obs}}{\partial x_\alpha} C^{-1} 
\frac{\partial \vec{S}^{\rm obs}}{\partial x_\beta},
\ea
where $x_\alpha$ indicates the parameters for our interest, i.e., $D_A$, 
$H^{-1}$, and so on. The quantity $\vec{S}^{\rm obs}$ is the 
signal vector expressed as 
\ba
\vec{S}^{\rm obs}=
\left(
\begin{array}{c}
\tilde{P}^{\rm obs}(\vec{k})
\\
\tilde{B}^{\rm obs}(\vec{k}_1,\vec{k}_2,\vec{k}_3)
\end{array}
\right)
\label{eq:vec_S_obs}
\ea
with $P^{\rm obs}$ and $B^{\rm obs}$ being given by Eqs.~(\ref{eq:Pk_obs}) 
and (\ref{eq:Bk_obs}). 
Note that in the absence of the cross covariance $C_{\rm PB}$, 
Eq.~(\ref{eq:fullfisher}) is reduced to a simplified form: 
\ba\label{eq:fisher_diag}
&&F_{\alpha\beta}\to \sum_{\vec{k}}
\frac{\partial \tilde{P}^{\rm obs}(\vec{k})}{\partial x_{\alpha}}C_{PP}^{-1}\frac{\partial \tilde{P}^{\rm obs}(\vec{k})}{\partial x_{\beta}} 
\\
&&\quad+\sum_{\vec{k}_1,\vec{k}_2,\vec{k}_3}
\frac{\partial \tilde{B}^{\rm obs}(\vec{k}_1,\vec{k}_2,\vec{k}_3)}{\delta x_{\alpha}}C_{BB}^{-1}\frac{\partial \tilde{B}^{\rm obs}(\vec{k}_1,\vec{k}_2,\vec{k}_3)}{\partial x_{\beta}}\nn 
\ea

Below, we will present the results in four different cases; 
1) power spectrum only, 2) bispectrum only, 
3)  power spectrum and bispectrum, but cross covariance ignored
[Eq.~(\ref{eq:fisher_diag})], and 
4) full information taking account of the non-vanishing cross covariance  
[Eq.~(\ref{eq:fullfisher})].

\begin{figure}
\begin{center}
\resizebox{3.4in}{!}{\includegraphics{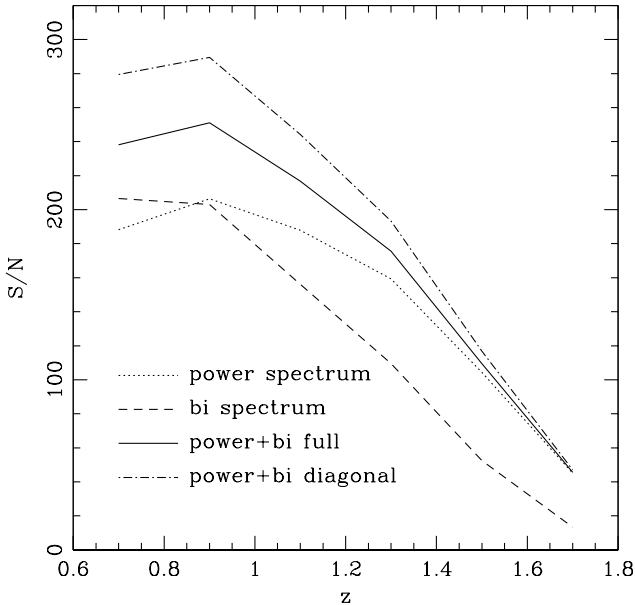}}
\end{center}
\caption{Signal-to-noise of power spectrum (dotted) and bispecrum (dashed) as function of redshift. Assuming a DESI-like experiment, the signal-to-noise ratios defined at Eqs.~(\ref{eq:SNR_pk}) and (\ref{eq:SNR_bk}) are estimated at each redshift bin. The combined result of the power spectrum and bispectrum, taking account of the cross covariance, is plotted as solid curve [see Eq.~(\ref{eq:SNR_pk_bk}) for definition], and, taking account of the diagonals only, is plotted as dot-dashed curve.
\label{fig:sofn}}
\end{figure}

\section{Results}
\label{sec:results}

In this section, we present the results of the Fisher matrix forecast described 
in the previous section. In what follows, the cosmological parameters needed 
to compute the power spectrum and bispectrum are set to the 
Planck 2013 concordance model \cite{Planck:2013XVI}; $\Omega_m=0.32$, $\Omega_b=0.049$, $h=0.67$, $A_S=2.15\times 10^{-9}$ and $n_S=0.96$. In our Fisher matrix, the number of free parameters to be estimated is six, i.e., geometric distances $D_A$ and $H^{-1}$, growth of structure $f$, galaxy biasing $b$ and $b_2$, and FoG damping $\sigma_p$. While the fiducial values of the first three parameters are specified by the cosmological parameters of the underlying cosmological model, the last three parameters are nuisance parameters, for which we assume that the observed galaxy distribution faithfully traces 
the mass distribution. That is, the fiducial values for the galaxy bias $b$ and $b_2$ are respectively set to $1$ and $0$, and the fiducial value of $\sigma_p$ is just given by the linear theory prediction.


\subsection{Signal-to-noise ratio}

Before discussing the constraints on each parameter, we first look at the signal-to-noise ratio $S/N$, and check the potential power of the bispectrum relative to that of the power spectrum. We compute the $S/N$ for both the power spectrum and bispectrum at each redshift bin. The signal-to-noise ratio for the power spectrum and bispectrum, $(S/N)_{\rm P}$ and $(S/N)_{\rm B}$, are respectively defined as,
\ba
\left(\frac{S}{N}\right)_{\rm P}^2&=&
\sum_{\vec{k}}\frac{\tilde{P}^2(\vec{k})}{C_{\rm PP}}
\label{eq:SNR_pk}\\
\left(\frac{S}{N}\right)_{\rm B}^2&=&
\sum_{\vec{k}_1,\vec{k}_2,\vec{k}_3}\frac{\tilde{B}^2(\vec{k}_1,\vec{k}_2,\vec{k}_3)}
{C_{\rm BB}}\,.
\label{eq:SNR_bk}
\ea
Similarly, we define the signal-to-noise ratio for the combined case:
\ba
\left(\frac{S}{N}\right)_{\rm P+B}^2&=&
\sum_{\vec{k}}\sum_{\vec{k}_1,\vec{k}_2,\vec{k}_3} 
~^{t}\vec{S}\, C^{-1}\, \vec{S}. 
\label{eq:SNR_pk_bk}
\ea
Here, the vector $\vec{S}$ is similarly defined as Eq.~(\ref{eq:vec_S_obs}), 
but the quantities $\tilde{P}^{\rm obs}$ and $\tilde{B}^{\rm obs}$ are replaced 
with $\tilde{P}$ and $\tilde{B}$ 
[Eqs.~\ref{eq:fullpks} and~\ref{eq:bifull}].

In Fig.~\ref{fig:sofn}, we plot three different $(S/N)$ defined 
above and one $(S/N)$ without the full covariance combination, adopting the specific survey design of DESI. 
Note that in all cases, the cut--off wavenumber $k$ is set to be $k=0.1\hompc$. 
The dotted and dash curves represent the S/N for power spectrum and bispectrum, respectively. The signal--to--noise for bispectrum is basically smaller than 
that for power spectrum. Nevertheless, the combination of the bispectrum with 
the power spectrum helps to improve the signal-to-noise ratio, as shown by the 
solid curve, and the improvement becomes significant at lower redshift. 
This is because the number density of galaxies becomes larger at lower 
redshift in our setup (Table \ref{tab:DESI}). Then, the shot noise 
contribution can be suppressed at relatively larger wavenumber. The combined $(S/N)$ with diagonals  only is presented as a dot-dashed curve.
Given the fact that the number of available configurations or modes in the 
bispectrum more rapidly increases with the 
wavenumber than that in power spectrum, 
$(S/N)_{\rm B}$ eventually exceeds $(S/N)_{\rm P}$ at 
the first redshift bin, although the cosmic variance error now 
comes to play an important role and the total signal-to-noise is 
slightly reduced compared to the one at the second redshift bin.

\subsection{Constraints on the geometric distances}

As we mentioned in Sec.~\ref{sec:intro}, a precision measurement of the BAO scale is the key to determining the geometric distances, $D_A(z)$ and $H^{-1}(z)$, through the A-P effect. Although the acoustic structure imprinted on the power spectrum and bispectrum actually depends on cosmology, the counterpart of the acoustic oscillations is precisely observed in the CMB anisotropies, and with the cosmological results by WMAP and Planck as prior information, the BAO can be used as standard ruler. At $k\lesssim 0.1\hompc$, the galaxy power spectrum and bispectrum are supposed to be described by the leading-order perturbation, where the acoustic signature is clearly visible. In the presence of galaxy bias, the only uncertainty is the clustering amplitude, however, this does not seriously affect the measurement of the characteristic scales of BAO.

\begin{figure}
\begin{center}
\resizebox{3.4in}{!}{\includegraphics{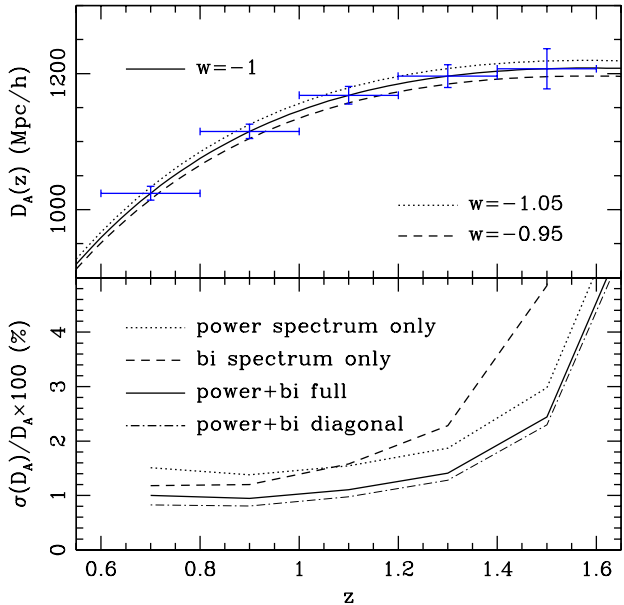}}
\end{center}
\caption{
{\it Upper}:  The expected $1\sigma$ error of the angular diameter distance $D_A(z)$ for DESI-like experiment. The size of the errorbars corresponds to the marginalized $1\sigma$ error estimated from the Fisher matrix, combining both the power spectrum and bispectrum. Dotted and dash curves represent $D_A(z)$ for different dark energy models with equations of state parameter $w=-1.05$ and $-0.95$ respectively. {\it Lower}: The fractional errors of $D_A$ using DESI are presented. Meaning of the line types are the same as in Fig.~\ref{fig:sofn}.
\label{fig:DA}}
\end{figure}

In Fig.~\ref{fig:DA},  the forecast result for the statistical errors of the angular diameter distance $D_A$ is shown. The fractional errors are presented in the lower panel. With the DESI-like survey, constraints on $D_A$ from the power spectrum (dotted) and bispectrum (dashed) become comparable at lower redshift, $z\lesssim1.2$, and precision is typically $1-2$\% level, although the constraining power of the bispectrum becomes rapidly worsen as increasing the redshifts. Note that 
if all the other parameters such as growth rate $f$ and galaxy bias are 
known a priori, $D_A$ can be measured with sub--percentage accuracy. However, 
the growth rate and the linear bias are mostly degenerate with $D_A$, particularly in the limit of $\mu\rightarrow 0$. Combining both the power spectrum and bispectrum, depicted as dot-dashed line, the constraint on $D_A$ will be improved by a factor of 1.5, thus mostly achieving a sub-percent precision. In the upper panel, we plot several curves for $D_A$ varying the dark energy equation of state parameter: $w=-1.05$ and $-0.95$ represented by dotted and dash curves respectively. The dark energy model will be severely constrained with the precision measurements of $D_A$.

Note that a comparable result of the constrained values from 
the power spectrum and bispectrum 
is understood from the signal--to--noise in Fig.~\ref{fig:sofn}. 
The signal--to--noise for bispectrum is higher than that of the power spectrum at low redshift bins, at which the number of observed galaxy is higher than the critical density of $n_{\rm gal}=10^{-3}\,(h^3{\rm Mpc}^{-3})$ (see Table~\ref{tab:DESI}). 
Accordingly, the cosmological constraints on distances from bispectrum become better than those from power spectrum in the first two redshift bins, 
as shown in Fig.~\ref{fig:DA}. Even if $n_{\rm gal}$ decreases but still exceeds $10^{-4}\,(h^3{\rm Mpc}^{-3})$, the constraints from bispectrum are comparable to those from power spectrum. Thus, combining both power spectrum and bispectrum 
provides an opportunity to improve the constraints on $D_A$. 

\begin{figure}
\begin{center}
\resizebox{3.4in}{!}{\includegraphics{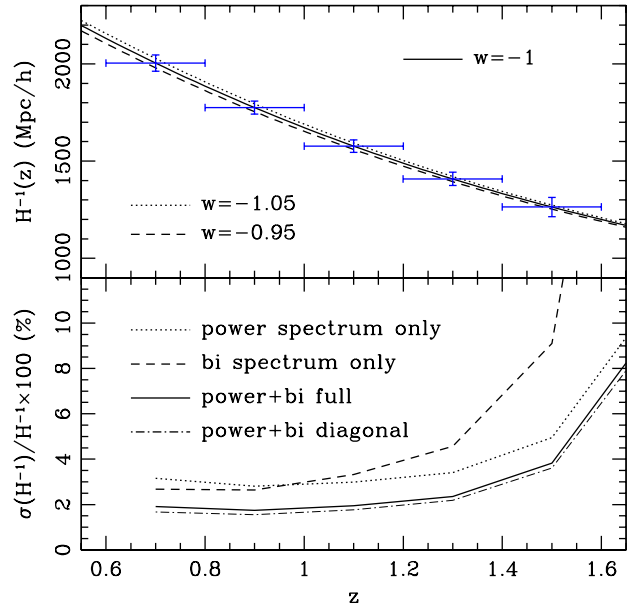}}
\end{center}
\caption{
{\it Upper}: Expected $1\sigma$ errors on the comoving radial distance, given by the 
inverse Hubble parameter $H^{-1}(z)$, assuming DESI-like survey. The results 
combining both the power spectrum and bispectrum are shown. Dotted and dash curves represent $H^{-1}(z)$ for different dark energy models with equations of state parameter $w=-1.05$ and $-0.95$ respectively. 
{\it Lower}: The fractional errors of $H^{-1}$ using DESI are presented. Meaning of the line types are the same as in Fig.~\ref{fig:sofn}.
\label{fig:Hinv}}
\end{figure}

On the other hand, the determination of radial distance $H^{-1}$ is more difficult than that of $D_A$. The redshift dependence of the error on $H^{-1}$ is presented in Fig.~\ref{fig:Hinv}. The dotted and dashed curves represent the constraints using the power spectrum and bispectrum, respectively. The information on $H^{-1}$ is most imprinted at the limit of $\mu\rightarrow 1$, at which the observed spectrum is influenced by RSD, mainly due to the non--linear smearing effect. Because of this, the resultant statistical precision is poorer than that of $D_A$ by a factor 2. Comparing between the solid and the dot-dashed curves, the off--diagonal elements of the covariance matrix do not significantly reduce the constraining power. For refrence, in the upper panel, we also plot $H^{-1}$ for various dark energy models with $w=-1.05$ and $-0.95$ represented by dotted and dash curves respectively.

\begin{figure}
\begin{center}
\resizebox{3.4in}{!}{\includegraphics{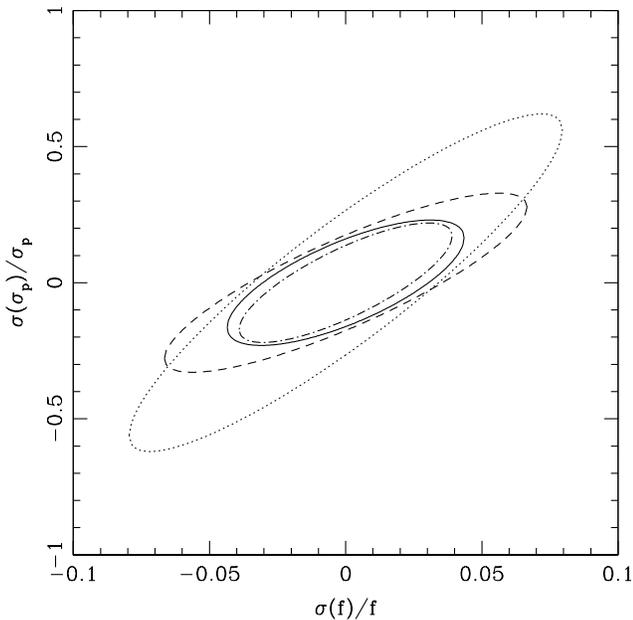}}
\end{center}
\caption{
Two-dimensional contour of the expected $1\sigma$ constraint on 
the parameters $f$ and $\sigma_p$. The results at the redshift bin of $0.6<z<0.8$ is particularly shown. The dotted contour represents the case of power spectrum, the dash contour represents the case of bispectrum, the dot-dashed contour represents the case of combination with diagonals only, and the solid curve represents the case of the full covariance combination.
\label{fig:confsigp}
}
\end{figure}

\subsection{Constraints on the growth of structure} 

At the scales close to the linear regime, the improved theoretical model of RSD successfully describes the observed spectrum. Still, however, nonlinear cross talk of the small-scale physics with large-scale anisotropic clustering (i.e., FoG effect) is significant, and a proper treatment of this cross talk needs to be incorporated. Here, we adopt a simple Gaussian form to phenomenologically describe the FoG effect [see Eqs.~(\ref{eq:FoG_power}) and~(\ref{eq:FoG_bispec})]. This would certainly remedy the flaw in the perturbative modeling of the RSD, however the strength of the damping characterized by $\sigma_p$ is now a free parameter, which needs to be determined by observations.

Fig.~\ref{fig:confsigp} shows the impact of this uncertainty in 
constraining the growth of structure. Here, the plotted result is the 
two-dimensional contour of the expected $1\sigma$ constraint, which is 
estimated at the specific redshift bin, $0.6<z<0.8$. 
The parameter $\sigma_p$ is significantly 
correlated with the linear growth rate $f$, and the analysis using 
the power spectrum alone, depicted as dotted contour, 
exhibits a strong parameter degeneracy. This is partly 
due to the small cutoff wavenumber, $k_{\rm max}=0.1\,h$\,Mpc$^{-1}$,  
below which the damping term, expressed as function of $(k\mu)^2$, is 
monotonically varied along the direction $\mu$, and the 
behavior looks very similar when we vary $f$. The situation almost remains 
unchanged even if we use the bispectrum, and the 
strong degeneracy between $f$ and $\sigma_p$ 
is observed (dashed contour). Interestingly, however, 
the direction of the 
parameter degeneracy differs from that of the power spectrum 
case. Accordingly, the combination of the power spectrum and bispectrum 
improves the constraint on $f$. The dot-dashed contour represents the case combination with diagonals only.
  
\begin{figure}
\begin{center}
\resizebox{3.4in}{!}{\includegraphics{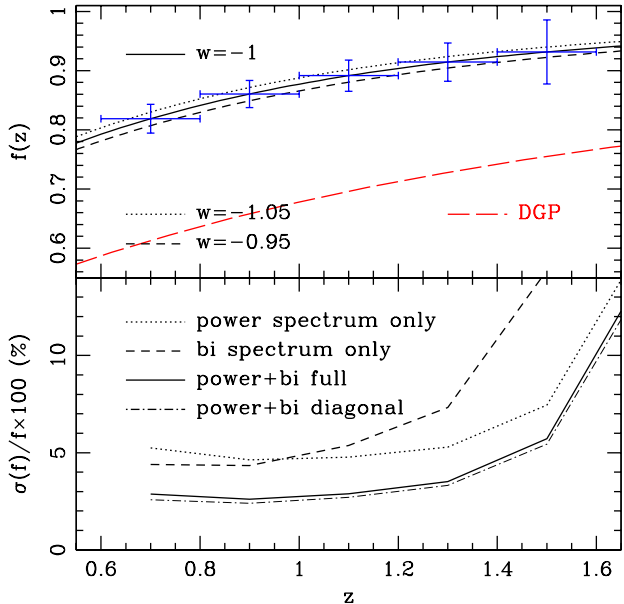}}
\end{center}
\caption{
{\it Upper}: Expected 1$\sigma$ constraint on the linear growth rate $f(z)$, assuming DESI.  The results are estimated using both the power spectrum and bispectrum. Dotted and dash curves represent $f(z)$ with $w=-1.05$ and $-0.95$ respectively, and red long dash curve represents $f(z)$ with DGP model.{\it Lower}: The fractional errors of $f$. Meaning of the line types are the same as in Fig.~\ref{fig:sofn}.
\label{fig:f}}
\end{figure}

\begin{figure}
\begin{center}
\resizebox{3.4in}{!}{\includegraphics{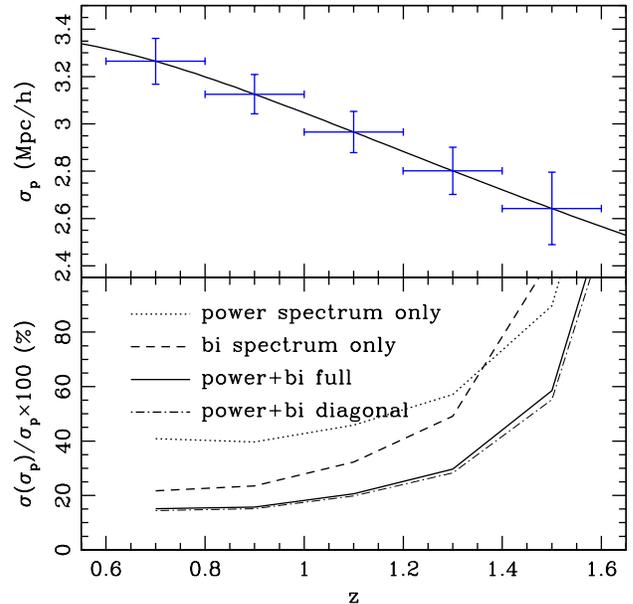}}
\end{center}
\caption{
{\it Upper}: The expected error on nuisance parameter 
characterizing the damping scales of the FoG effect, $\sigma_p(z)$. 
The results are presented assuming DESI-like experiment.   
{\it Lower}: The fractional errors of $\sigma_p(z)$.
Meaning of the line types are the same as in Fig.~\ref{fig:sofn}. 
\label{fig:sigp}}
\end{figure}

Figs.~\ref{fig:f} and \ref{fig:sigp} respectively show the marginalized results of the fractional errors on $f$ and $\sigma_p$, plotted as a function of redshift. Combining power spectrum and bispectrum, the constraint on $f$ becomes tighter at all redshift bins. In particular, at lower redshift bins, statistical power of the bispectrum is enhanced, and the relative impact on the constraint on the parameter $\sigma_p$ eventually becomes stronger (see bottom panel of Fig.~\ref{fig:sigp}). As a result, compared to the power spectrum results, the combined result of the constraint on $f$ is improved by a factor of $2$. Note that similar to the constraints on the geometric distance, the effect of the off-diagonal components of the error covariance are insignificant in the case of $f$ and $\sigma_p$, 
and it does not appreciably change the results. 
In the upper panel, we plot $f$ for different theoretical models. The black dotted and dash curves represent $f$ for dark energy models with $w=-1.05$ and $-0.95$ respectively. It is interesting to note that we are able to constrain the dark energy from the measured $f$ alone. Also, the red long dash curve represents $f$ of DGP model~\cite{Dvali:2000hr,Song:2006jk,Koyama:2005kd}, demonstrating the outperformance of the stage-IV class survey.
  
\begin{figure}
\begin{center}
\resizebox{3.4in}{!}{\includegraphics{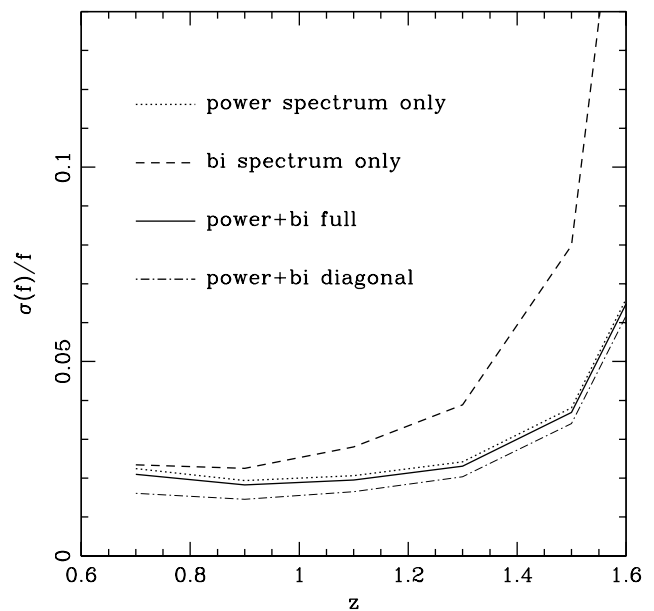}}
\end{center}
\caption{
Fractional errors of the linear growth rate $f$ in the case when we know a priori the FoG effect (i.e., $\sigma_p$). Meaning of the line types is the same as in Fig.~\ref{fig:sofn}.
\label{fig:fnoFoG}}
\end{figure}

Finally, to see the impact of the FoG effect on the measurement of the growth rate, 
we plot in Fig.~\ref{fig:fnoFoG} the marginalized constraints of $f$ in the 
case when we know the FoG effect a priori. As it is expected, the parameter 
$f$ is better constrained, and the precision reaches at $\sim2\%$ level. 
One notable and interesting point is that the power spectrum always 
gives a tighter constraint, and combination of power spectrum and bispectrum 
does not improve much the constraint. Thus, the benefit of 
using the bispectrum is substantially reduced.

\subsection{Constraints on the galaxy bias parameters}

The combination of power spectrum and bispectrum has been frequently used in the literature to constrain the bias parameters (e.g., \cite{Gil-Marin:2014sta,Gil-Marin:2014baa} for recent results). With the survey like DESI, the constraint on these parameter will become much more tigher. This is 
true even if we simultaneously estimate the expansion history and 
growth of structure.

\begin{figure}
\begin{center}
\resizebox{3.4in}{!}{\includegraphics{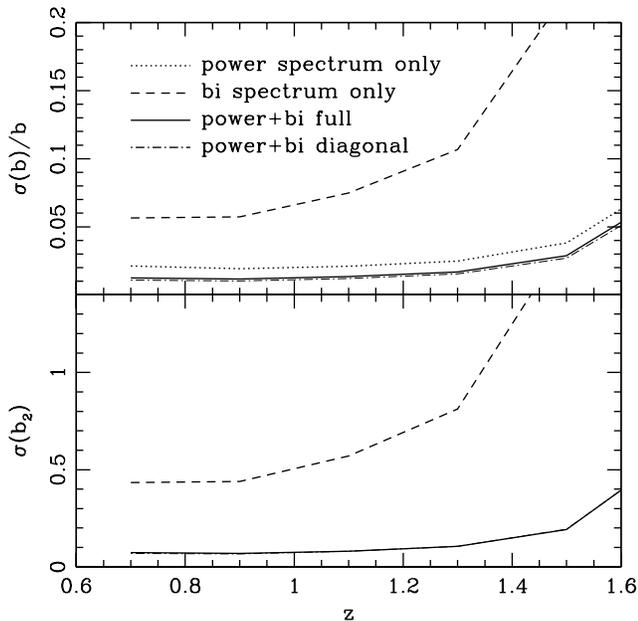}}
\end{center}
\caption{
Expected error on galaxy bias parameters. Upper panel shows the fractional errors on linear bias $b$, while the lower panel presents the errors on second-order bias $b_2$. Meaning of the line types is the same as in Fig.~\ref{fig:DA}.
\label{fig:bias}}
\end{figure}

Fig.~\ref{fig:bias} shows  
the forecast result of the statistical errors on galaxy bias parameters. 
The upper panel presents the constraint on the linear bias parameter $b$. 
The linear bias is tightly constrained by the power spectrum depicted as dotted 
line, and no benefit to use the bispectrum is found (dashed), 
although the combination of power spectrum and bispectrum still improves the constraint by a factor of 2 (solid). This is because of 
another bias parameter $b_2$, which coherently boosts the amplitude of 
bispectrum. Also, one important assumption here is that the cosmological parameters are known a priori from the CMB observations, including the normalization of linear power spectrum.  In other words, 
if the specific cosmological models are not given, the linear bias information is not extracted from the power spectrum.

On the other hand, the bottom panel of Fig.~\ref{fig:bias} shows the 
constraint on the nonlinear bias parameter $b_2$. 
The power spectrum alone cannot constrain $b_2$, 
because the parameter $b_2$ does not appear in the expression of 
power spectrum, relevant at the large scales of our interest 
(see Sec.~\ref{subsec:power_spectrum}). Even using 
the bispectrum, $b_2$ is poorly constrained. This is due to the 
degeneracy between the bias parameters $b$ and $b_2$. 
In this respect, the combination of the power spectrum and bispectrum is 
quite essential to break the degeneracy. With the help of the power spectrum
information, 
$b$ is measured precisely, and this will lead to a substantial improvement of 
the constraint on $b_2$, as shown by the solid line.

\section{Summary}
\label{sec:summary}

The bispectrum has been recognized as a powerful tool to probe non--Gaussian initial conditions, and to enhance the constraints on BAO distance measurements. In this paper, we highlighted the benefit to improve the simultaneous constraint on the growth of structure, and geometric distances through the RSD and A-P effect. The power spectrum in redshift space suffers from nonlinear cross talk between small- and large-scale clustering, and the clustering amplitude is significantly suppressed at large scales.  While the impact of this FoG effect is
less significant at the scales close to the linear regime, it is difficult to break degeneracy between the distortions induced by coherent motions and systematics arising from the FoG effect. Although this would be certainly resolved with a better understanding of the nonlinear cross talk based on a more elaborate theoretical model, we proposed an alternative method to improve the cosmological constraints by using both the power spectrum and bispectrum.
Assuming a DESI-like experiment, Fisher matrix analysis suggests that 
the analysis using the power spectrum data alone shows a 
strong parameter degeneracy between the growth of structure ($f$) and $\sigma_p$ as a nuisance parameter characterizing the damping scale of the FoG effect. 
In hat case the estimated error on $f$ is about 5$\%$ at the most sensitive redshift bin. 
We found that this  
parameter degeneracy can be broken when combining both the power spectrum and 
bispectrum. As a result, the constraint will be improved by a factor of two, which satisfies the primary science goal to distinguish the possibilities to explain the cosmic acceleration, i.e., the dark energy and modification of gravity on cosmological scales. 

Note finally that these results certainly depend on the model of the FoG effect. This is especially the case for the bispectrum. While we adopted a simple Gaussian ansatz to describe the FoG damping, a proper way to characterize the FoG effect in bispectrum is not yet fully understood. 
Different assumption or prescription of the FoG effect is possible, and may lead to a quantitatively different result. We leave this issue for further work.  

\acknowledgements{Numerical calculations were performed by using a high performance computing cluster in the Korea Astronomy and Space Science Institute. This work is supported in part by a Grant-in-Aid for Scientific Research from the Japan Society for the Promotion of Science (No.~24540257).}

\end{document}